
\input epsf                                                               %
\input harvmac
\def\Title#1#2{\rightline{#1}\ifx\answ\bigans\nopagenumbers\pageno0\vskip1in
\else\pageno1\vskip.8in\fi \centerline{\titlefont #2}\vskip .5in}

%
%
\ifx\epsfbox\UnDeFiNeD\message{(NO epsf.tex, FIGURES WILL BE IGNORED)}
\def\figin#1{\vskip2in}
\else\message{(FIGURES WILL BE INCLUDED)}\def\figin#1{#1}
\fi
\def\Fig#1{Fig.~\the\figno\xdef#1{Fig.~\the\figno}\global\advance\figno
 by1}
%
%
%
%
\def\ifig#1#2#3#4{
\goodbreak\midinsert
\figin{\centerline{\epsfysize=#4truein\epsfbox{#3}}}
\narrower\narrower\noindent{\footnotefont
{\bf #1:}  #2\par}
\endinsert
}
\def\ifigx#1#2#3#4{
\goodbreak\midinsert
\figin{\centerline{\epsfxsize=#4truein\epsfbox{#3}}}
\narrower\narrower\noindent{\footnotefont
{\bf #1:}  #2\par}
\endinsert
}
%
%
%
\font\ticp=cmcsc10

\def\ajou#1&#2(#3){\ \sl#1\bf#2\rm(19#3)}
\def\jou#1&#2(#3){\unskip, \sl#1\bf#2\rm(19#3)}
\def\RN{Reissner-Nordstrom}
\def\rst{r_*}
\def\Mpl{M_{\rm pl}}
\def\threeg{{}^3g}

\def\cald{{\cal D}}

%
%
\lref\SuUg{L. Susskind and J. Uglum, ``Black hole entropy in canonical
quantum gravity and superstring theory,'' hep-th/9401070
\jou Phys. Rev. &D50 (94) 2700.}
\lref\FPST{T.M. Fiola, J. Preskill, A. Strominger, and S.P. Trivedi,
``Black hole  thermodynamics and information loss in two-dimensions,''
hep-th/9403137 \jou Phys. Rev. &D50 (94) 3987.}
\lref\DXBH{S.B. Giddings and A. Strominger, ``Dynamics of extremal black
holes,'' hep-th/9202004\jou Phys. Rev. &D46 (92) 627.}
\lref\CaTi{S. Carlip and C. Teitelboim, ``The off-shell black hole,''
gr-qc/9312002, IAS/UC Davis preprint IASSNS-HEP-93/84=UCD-93-34.}
\lref\herring{S. Coleman, ``Black Holes as Red Herrings:
Topological Fluctuations and the Loss of Quantum Coherence,''\ajou Nucl.
Phys. & B307 (88) 864.}
\lref\LInc{S.B. Giddings and A. Strominger, ``Loss
of Incoherence and Determination of Coupling Constants
in Quantum Gravity,"\ajou Nucl. Phys. &B307 (88) 854.}
\lref\ernst{F. J. Ernst, \ajou J. Math. Phys. &17 (76) 515.}
\lref\CGHS{C.G. Callan, S.B. Giddings, J.A. Harvey and A. Strominger,
``Evanescent black holes,'' hep-th/9111056 \jou Phys. Rev. &D45 (92) R1005.}
\lref\Trieste{S.B. Giddings, ``Quantum mechanics of black holes,''
hep-th/9412138, UCSBTH-94-52 (1994
Trieste lectures).}
\lref\HaSt{J.A. Harvey and A. Strominger, ``Quantum aspects of black
holes,'' hep-th/9209055,
in proceedings of the 1992 Trieste Spring School on String
Theory and Quantum Gravity.}
\lref\Pres{J. Preskill, ``Do black holes destroy information?''
hep-th/9209058, in the
proceedings of the International Symposium on Black holes, Membranes,
Wormholes and Superstrings,
Woodlands, TX, 16-18 Jan 1992.}
\lref\Japan{S.B. Giddings, ``Black holes and quantum predictability,''
hep-th/9306041, in {\sl Quantum
Gravity},  (Proceedings of the 7th
Nishinomiya Yukawa Memorial Symposium), K. Kikkawa and M. Ninomiya, eds.
(World Scientific, 1993).}
\lref\BDDO{T. Banks, A. Dabholkar, M.R. Douglas, and M O'Loughlin, ``Are
horned particles the climax of Hawking evaporation?'' hep-th/9201061
\jou Phys. Rev. &D45 (92) 3607.}
\lref\BaOl{T. Banks and M. O'Loughlin, ``Classical and quantum production
of cornucopions at energies below $10^{18}$ GeV,'' hep-th/9206055
\jou Phys.Rev. &D47 (93) 540-553.}
\lref\BOS{T. Banks, M. O'Loughlin, and A. Strominger, ``Black hole remnants
and the information puzzle,'' hep-th/9211030 \jou Phys. Rev.& D47 (93) 4476.}
\lref\CILAR{S.B. Giddings, ``Comments on information loss and remnants,''
 hep-th/9310101 \jou Phys. Rev. & D49 (94) 4078.}
\lref\Gibbsp{G.W. Gibbons, ``Vacuum polarization and spontaneous loss of
charge by black holes,''\ajou Comm. Math. Phys. &44 (75) 245.}
\lref\Gibb{G.W. Gibbons, ``Quantized flux tubes in Einstein-Maxwell
theory and noncompact internal
spaces,'' in {\sl Fields and geometry}, proceedings of
22nd Karpacz Winter School of Theoretical Physics: Fields and
Geometry, Karpacz, Poland, Feb 17 - Mar 1, 1986, ed. A. Jadczyk (World
Scientific, 1986).}
\lref\GaSt{D. Garfinkle and A. Strominger, ``Semiclassical Wheeler wormhole
production,''\ajou Phys. Lett. &B256 (91) 146.}
\lref\DGKT{H.F. Dowker, J.P. Gauntlett, D.A. Kastor and J. Traschen,
``Pair creation of dilaton black holes," hep-th/9309075
\jou Phys. Rev. & D49 (94) 2209.}
\lref\DGGH{F. Dowker, J. Gauntlett, S.B. Giddings, and G.T. Horowitz,
``On pair creation of extremal black holes and Kaluza-Klein
monopoles,'' hep-th/9312172 \jou Phys. Rev. &D50 (94) 2662.}
\lref\CaWi{R.D. Carlitz and R.S. Willey, ``Reflections on moving
mirrors,''\ajou Phys. Rev. &D36 (87) 2327; ``Lifetime of a black
hole,''\ajou Phys. Rev. &D36 (87) 2336.}
\lref\PoSt{J. Polchinski and A. Strominger, ``A possible resolution of the
black hole information puzzle,'' hep-th/9407008, UCSB preprint
UCSB-TH-94-20.}
\lref\Stro{A. Strominger, ``Unitary rules for black hole evaporation,''
hep-th/9410187, UCSB preprint UCSBTH-94-34.}
\lref\Beke{J.D. Bekenstein, ``Do we understand black hole entropy?''
gr-qc/9409015, talk at Seventh Marcel Grossman meeting, Stanford Univ.}
\lref\StTr{A. Strominger and S. Trivedi, ``Information consumption by
Reissner-Nordstrom black holes,'' hep-th/9302080
\jou Phys. Rev. & D48 (93) 5778.}
\lref\GiHa{G.W. Gibbons and S.W. Hawking, ``Action integrals and partition
functions in quantum gravity,''\ajou Phys. Rev. &D15 (77) 2752.}
\lref\tunneling{H.J. DeVega, J.L. Gervais, and B. Sakita, ``Real time
approach to instanton phenomena (II): multidimensional potential with
continuous symmetry,''\ajou Nucl. Phys. & B142 (78) 125\semi
I. Bender and H.J. Rothe, ``Tunneling potentials and WKB wavefunctionals
along multi-instanton paths in an SU(2) gauge theory,''\ajou Nucl. Phys.
&B142 (78) 177, and references therein.}
\lref\BiCh{K. Bitar and S.-J. Chang, ``Vacuum tunneling and fluctuations
around a most probable escape path,''\ajou Phys. Rev. & D18 (78) 435.}
\lref\CoUses{S. Coleman, ``The uses of instantons,'' in S. Coleman,
{\sl Aspects of
Symmetry: selected Erice lectures} (Cambridge Univ. Press, 1985). }
\lref\GGS{D. Garfinkle, S.B. Giddings, and A. Strominger, ``Entropy in
Black Hole Pair Production,'' gr-qc/9306023 \jou Phys. Rev. &D49 (94) 958.}
\lref\Melv{M. A. Melvin ``Pure magnetic and electric
geons,''\ajou Phys. Lett. &8 (64) 65.}
\lref\Ernst{F. J. Ernst, ``Removal of the nodal singularity of the
C-metric,'' \ajou J. Math. Phys. &17 (76) 515.}
\lref\Triv{S.P. Trivedi, ``Semiclassical extremal black holes,''
hep-th/9211011 \jou Phys. Rev. &D47 (93) 4233.}
\lref\BGHS{B. Birnir, S.B. Giddings, J.A. Harvey and A. Strominger,
``Quantum black holes,'' hep-th/9203042 \jou Phys. Rev. &D46 (92) 638.}
\lref\SuTh{L. Susskind and L. Thorlacius, ``Hawking radiation and
back-reaction,'' hep-th/9203054, \jou Nucl. Phys. &B382 (92) 123.}
\lref\Yi{P. Yi, ``Toward one-loop tunneling rates of near-extremal magnetic
black hole pair-production,'' hep-th/9407173, Caltech preprint CALT-68-1936.}
\lref\LNW{K. Lee, V.P. Nair, and E.J. Weinberg, ``A classical instability
of Reissner-Nordstrom solutions and the fate of
magnetically charged black holes,'' hep-th/9111045
\jou Phys. Rev. Lett. &68 (92) 1100.}
\lref\CBHR{S.B. Giddings, ``Constraints on black hole remnants,''
hep-th/930402 \jou Phys. Rev. &D49 (94) 947.}
\lref\BPS{T. Banks, M.E. Peskin, and L. Susskind, ``Difficulties for the
evolution of pure states into mixed states,''\ajou Nucl. Phys. &B244 (84)
125.}
\lref\Sred{M. Srednicki, ``Is purity eternal?,'' hep-th/920605 \jou
Nucl. Phys. &B410 (93) 143.}
\lref\tHoo{G. 't Hooft, ``Dimensional reduction in quantum gravity,''
gr-qc/9310006, Utrecht preprint THU-93/26, and references therein.}
\lref\STU{L. Susskind, L. Thorlacius, and J. Uglum, ``The stretched horizon
and black hole complementarity,'' hep-th/9306069
\jou Phys. Rev. &D48 (93) 3743.}
\lref\Suss{L. Susskind, ``The world as a hologram,'' hep-th/9409089,
Stanford preprint
SU-ITP-94-33, and references therein.}
\lref\FPST{T.M. Fiola, J. Preskill, A. Strominger, and S. Trivedi,
``Black hole thermodynamics and information loss in two-dimensions,''
hep-th/9403137 \jou Phys. Rev. &D50 (94) 3987.}
\lref\Thor{L. Thorlacius, ``Black hole evolution,'' hep-th/9411020,
NSF-ITP-94-109 (lectures at 1994 Trieste Spring School).}

\Title{\vbox{\baselineskip12pt\hbox{UCSBTH-94-50}
\hbox{hep-th/9412159}
}}
{\vbox{\centerline {Why Aren't Black Holes Infinitely Produced?}
}}

\centerline{{\ticp Steven B. Giddings}\footnote{$^\dagger$}
{Email addresses:
giddings@denali.physics.ucsb.edu} }
\vskip.1in
\centerline{\sl Department of Physics}
\centerline{\sl University of California}
\centerline{\sl Santa Barbara, CA 93106-9530}
\bigskip
\centerline{\bf Abstract}

Unitarity and locality imply a remnant solution to the information problem,
and also imply that \RN\ black holes have infinite numbers of internal
states.  Pair production of such black holes is reexamined including the
contribution of these states.  It is argued that the rate is proportional
to the thermodynamic quantity
$Tr e^{-\beta H}$, where the trace is over the internal
states of a black hole; this is in agreement with estimates from an
effective field theory for black holes.  This quantity, and the rate,
is apparently infinite due to the infinite number of states.
One obvious out is if the number of internal states of a black hole is
finite.

\Date{}

\newsec{Introduction}

Despite much recent effort the problem of what happens to
quantum-mechanical information thrown into a black hole remains a puzzling
problem.  A variety of detailed scenarios can be boiled down to three basic
pictures:\foot{For reviews see \refs{\HaSt\Pres\Japan\Thor-\Trieste}.}
information is destroyed,
information is returned in the Hawking radiation, or information is left
behind in a black hole remnant.
As is by now well known, if one attempts to describe black hole formation
and evaporation from a low-energy, effective point of view each of these
possibilities encounters serious conflicts with basic low energy principles
such as energy conservation, locality, and crossing symmetry.

Those advocating a remnant scenario\refs{\CGHS\BDDO-\DXBH} have attempted evade
the problem of infinite production by hypothesizing that black hole
remnants are not correctly described by low-energy effective field theory
and/or that crossing somehow fails\refs{\BaOl\BOS-\CILAR}.  Fertile ground
for the investigation of these possibilities is provided by the phenomenon
of pair-production of charged black holes in background electromagnetic
fields.  If one hypothesizes that information is neither destroyed nor
re-emitted, then there should be an infinite number of internal states of
such a black hole:  one can feed in arbitrarily large amounts of
information-rich matter, then allow evaporation to extremality\StTr.
Furthermore, instantons for such processes are described in
\refs{\Gibb\GaSt\DGKT-\DGGH}.  Remnant advocates have hoped that a
reliable calculation of the resulting rate for pair creation would be
finite, and that charged black holes would therefore serve as a guide to
formulation of theories of infinitely-degenerate remnants with finite
production, which might also extend to neutral remnants.

Indeed, this testing ground is critical.  Charged black holes
might show us the way to a theory of remnants, but if they do not, then the
remnant hypothesis apparently
dies with them.  The reason is that the basic postulates
of the remnant hypothesis, namely unitarity (no information destruction)
and locality (no reradiation of information in Hawking information)
imply an infinite degeneracy of charged black holes, and if
this implies that charged black holes are infinitely pair-produced then
these postulates are not correct, removing the raison d' \^etre for neutral
remnants. Pair production of charged black holes
is thus a litmus test for the theory of remnants.

This paper will at the outset accept these postulates and attempt to
investigate their viability through a more careful investigation of the
pair production problem for \RN\ black holes.  The essential features of
these arguments extend as well to pair production of dilatonic black
holes\refs{\DGKT,\DGGH}.  It begins by reviewing some of the basic features
of remnant theories and the argument for infinite pair production that
follows from an effective description, as well as issues that remnants
raise for black hole thermodynamics.  Next the role of \RN\ black holes
as remnants of the Hawking process in the charged sector is reviewed, and
a description of the infinite states appropriate to an outside observer is
outlined.  The following section contains a reinvestigation of the
Schwinger process for \RN\ black holes.  It is argued that the contribution
of the infinite number of states is contained in the fluctuation
determinant around the instanton, and that this cannot be computed without
full knowledge of Planck scale physics.  However, the calculation is nearly
identical to that of Tr$e^{-\beta H}$ for a black hole in contact with a
heat bath, and if
there are an infinite number of nearly degenerate
black hole states, then this appears
infinite independent of our inability to
describe them explicitly.  The emergence of such a factor agrees with the
rate computation done in the effective approach.
In closing, possibilities for avoiding infinite production of \RN\ black
holes are outlined.

\newsec{Basics of remnants}

As stated in the introduction, the postulates of unitarity and locality
imply that the information lost in the Hawking evaporation of a black hole
is left behind in a black hole remnant.  Consider an initial black hole of
mass $M$ that leaves behind a remnant.  This should have a mass $m\sim
\Mpl$, and due to the difficulty in emitting its large information $I\sim
M^2$ with its
small available energy, it will have a very long lifetime\refs{\CaWi,\Pres},
\eqn\decayt{\tau\sim M^4\ .}
With an arbitrary initial black hole, an arbitrarily large amount of
information can be stored in such a remnant, and so there must be an
infinite number of internal states or species of such an object.

At first sight two obvious issues leap to mind.  First, the infinite number
of remnant species would appear to lead to infinite total remnant
production rates in various physical processes.  Second, it seems rather
strange to have absolutely stable remnants, and it is not obvious what
physics would give a remnant decay time as in \decayt.

\ifig{\Fig\One}{The Penrose diagram for an evaporating neutral black hole,
together with a time slicing.}{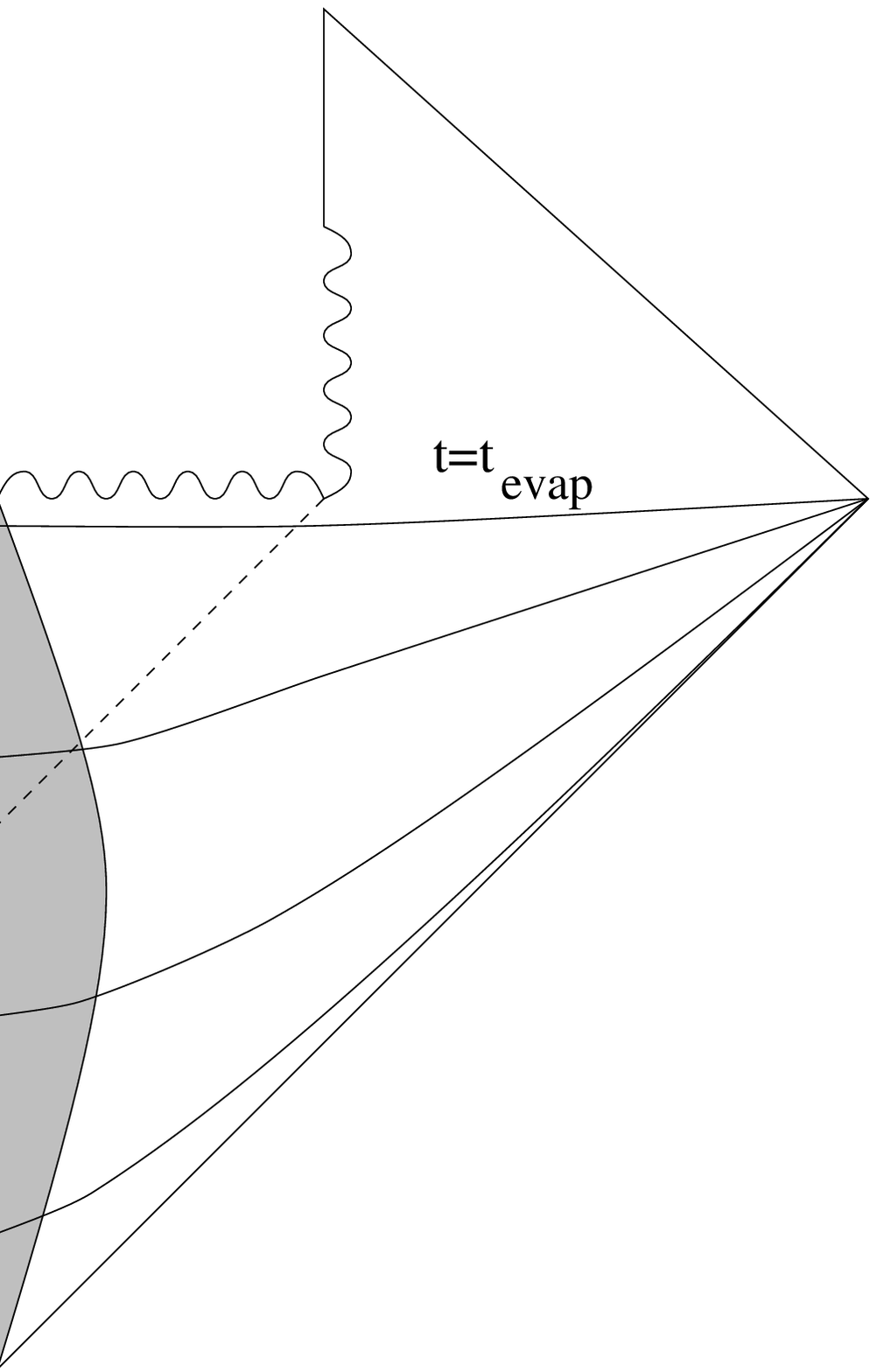}{3.0}

Answers to the second question may be provided by understanding how the
internal physics of remnants returns the information and
respects the constraints placed by the relationship between
information and energy.  One approach to this physics
has been recently proposed by Polchinski
and Strominger\refs{\PoSt,\Stro}.  They discuss the proper treatment of a
scenario where the black hole interior branches off a baby universe in
an attempt to carry information off.  As in the case of baby universes,
there is not a repeatable loss of
information\refs{\herring,\LInc,\Trieste}, and the couplings
self-adjust so that the interior takes a long time to split off and a
long-lived remnant results.  Alternatively, note that if we consider a late
time slice through a plausible Penrose diagram of an evaporating black
hole, figs.~1, 2, this slice consists of a planckian fiber attached to a
flat geometry.  With this picture of a remnant in mind, it is quite
plausible that the appropriate behavior follows from the necessarily
planckian physics of the fiber.  Thus solutions to the second problem are
easily imagined.

\ifig{\Fig\Two}{The spatial geometry of a late time slice through
fig.~1.  As the radius of the black hole decreases through evaporation,
the slice becomes a thin, and in the limit, planckian
fiber attached to the asymptotic geometry.}{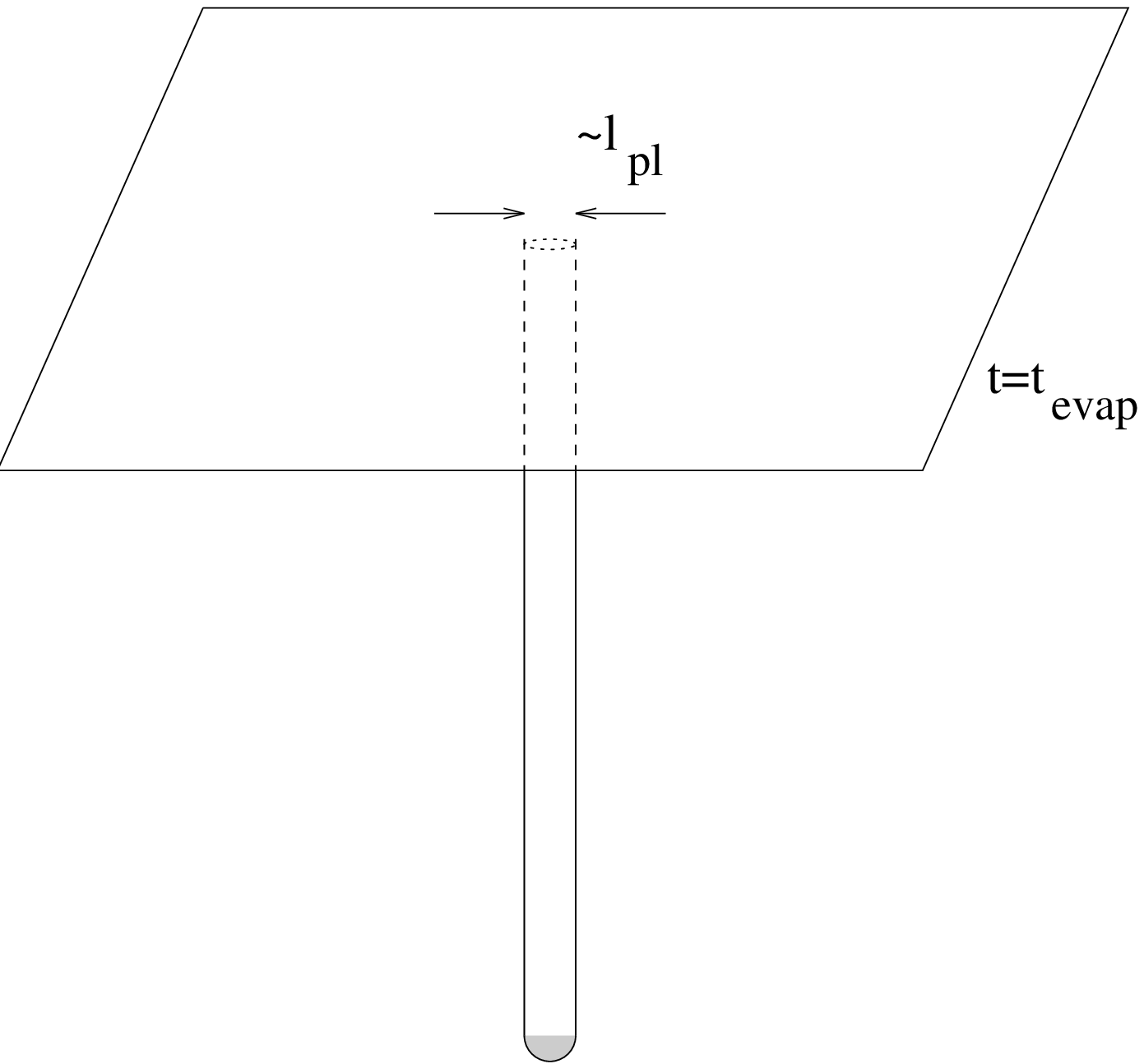}{2.0}

The first issue is much more difficult.  Since remnants are localized
massive objects, we expect that the only description of them that is
local/causal, Lorentz invariant, and quantum mechanical is in terms of an
effective field, $\phi_a$.  Here $a$ is a species label.
The couplings of the remnant field to other fields may
be quite complicated, but near zero momentum transfer their approximate
form would appear to be
dictated simply by the mass and charge of the remnant.  If for
simplicity we think of electrically
charged scalar remnants (the magnetic case follows via electromagnetic
duality), they should therefore be
described by an effective action
\eqn\remact{S_{\rm eff} = \int d^4 x \sum_a \left( -|D_\mu \phi_a|^2 -
m_a^2 |\phi_a|^2 \right) + \cdots\ ,}
with $ D_\mu = \partial_\mu +iQA_\mu $,  and where higher dimension terms are
not written.  At low momenta transfers
the latter terms are expected to be negligible.

Such a coupling will allow Schwinger production of pairs of remnants.  The
decay rate of a background electric field $A_\mu^0$ is given by the
imaginary part of the euclidean vacuum-to-vacuum amplitude,
\eqn\drate{ V_4 \Gamma = 2 {\rm Im\ ln} \left( \int \cald {A}_\mu
 \cald \phi e^{-S[A^0+A]-S_{\rm eff}[\phi_a]} \bigg/ \int \cald {A}
 \cald \phi e^{-S[A]-S_{\rm eff}[\phi_a]} \right)}
where $V_4$ is the four-volume and  $S[A]$ is the Maxwell action.  To
lowest order in the coupling electromagnetic fluctuations are neglected,
and one finds
\eqn\dratea{V_4 \Gamma = {\rm Im} \ln {\rm det} \left\{\left[
 -\left( \partial_\mu + iQA_\mu^0 \right)^2 + m_a^2 \right]
\bigg/ \left[
 -\partial_\mu^2 + m_a^2 \right]\right\} \ .}
%
Then $ln\,det=Tr\,ln$, and the operator
traces can be rewritten in terms of single
particle amplitudes, giving
\eqn\decayr{\eqalign{
{\rm Im\, ln\, det}& \left[
 -\left(\partial_\mu + iQA_\mu^0 \right)^2 + m_a^2 \right]
= 2 {\rm Im} \int_0^\infty {dT\over T}
\int_{X(0)=X(T)} \cald X \cr &\exp\left\{ -\int_0^T d\tau \left(
{{\dot X}^2\over 2} + iQ A_\mu {\dot
X}^\mu\right)\right\} {\rm tr}_a \exp\left\{ -{T\over 2} m_a^2  \right\}\
. }}

Each term in the sum over $a$ is well approximated by a Schwinger saddlepoint
corresponding to circular euclidean motion, and the decay rate is then
given by
\eqn\Srate{\Gamma\sim \sum_a e^{-\pi m_a^2/QE}\ .}
If there are an infinite number of species, the sum
diverges and the total production rate is infinite.
If we furthermore suppose that the remnant spectrum consists of nearly
degenerate states, $m_a = M + \Delta m_a$, with $\Delta m \ll M$,
then \Srate\ can be rewritten
\eqn\Prate{\Gamma\sim e^{-\pi M^2/QE} {\rm Tr}_a e^{-\beta \Delta m}\ ,}
with $\beta=2\pi M/QE$.  Thus it is proportional to the partition function
for the nearly-degenerate states.

One might attempt to find remnant effective theories that avoid this
problem, perhaps by avoiding minimal couplings altogether.\foot{It is
conceivable that there are theories with no minimal couplings to remnants,
in which minimal couplings are mimicked by more complicated couplings.  For
example, it may be that there is no sense in which we can scatter a charged
particle off an extremal charged black hole without exciting it.}
However, no such theories have been formulated.  Furthermore, as will be
shown in sec.~4, a remarkably similar result holds for Schwinger
production of black holes.
%

In closing this general discussion, it is also worth emphasizing
the conflict between theories of remnants and black hole
thermodynamics\refs{\Beke}.
In particular, in a remnant scenario the Bekenstein-Hawking
entropy of a black hole is not
related in any obvious way to the number of its internal states.  The
Bekenstein bound is directly violated by remnants.
In a viable remnant scenario
it may be that the only role of black hole thermodynamics is to furnish a
macroscopic description of properties of the black hole: for example, the
temperature versus mass relation is independent of the number of black hole
internal states.

\newsec{Extremal \RN\ states}

In studying the possibility of a remnant solution to the information
problem it is useful to investigate
a situation where the Hawking process is
guaranteed to leave a remnant:  the evaporation of a
charged black hole.

Indeed, following the preceeding logic, suppose that we begin with a
charged\foot{This discussion will consider either electrically or
magnetically charged black holes; issues connected with Schwinger
charge loss of an electric black hole will be postponed to a
subsequent section.} black hole.  There are several ways that one
could be obtained; it could come from collapse of
charged matter, or be one end of a Wheeler wormhole created in
Schwinger production, or be an extremal black hole either of
primordial origin or created in the Schwinger process.  In each case
the global geometry of the solution differs.  However, each shares
the important feature that the endpoint of the Hawking process leaves
an infinite number of internal black hole states, and that these
states are practically indistinguishable independent of the origin of
the black hole.

To see this, notice that we can throw an arbitrary matter
configuration into our black hole over an arbitrarily long time.  We
then allow it to evaporate; in each case the endpoint of the
evaporation process is a black hole with the extremal value of the
mass, $M=Q$.  If the information from the infalling matter neither
escapes the black hole nor is annihilated, then we have managed to
construct an infinite variety of internal black hole quantum
states depending
on the configuration of the infalling matter.

\ifig{\Fig\Rloc}{Shown is the geometry motivating the definition of the
radiolocation coordinates $(r_*,t)$ of an arbitrary point
$P$.}{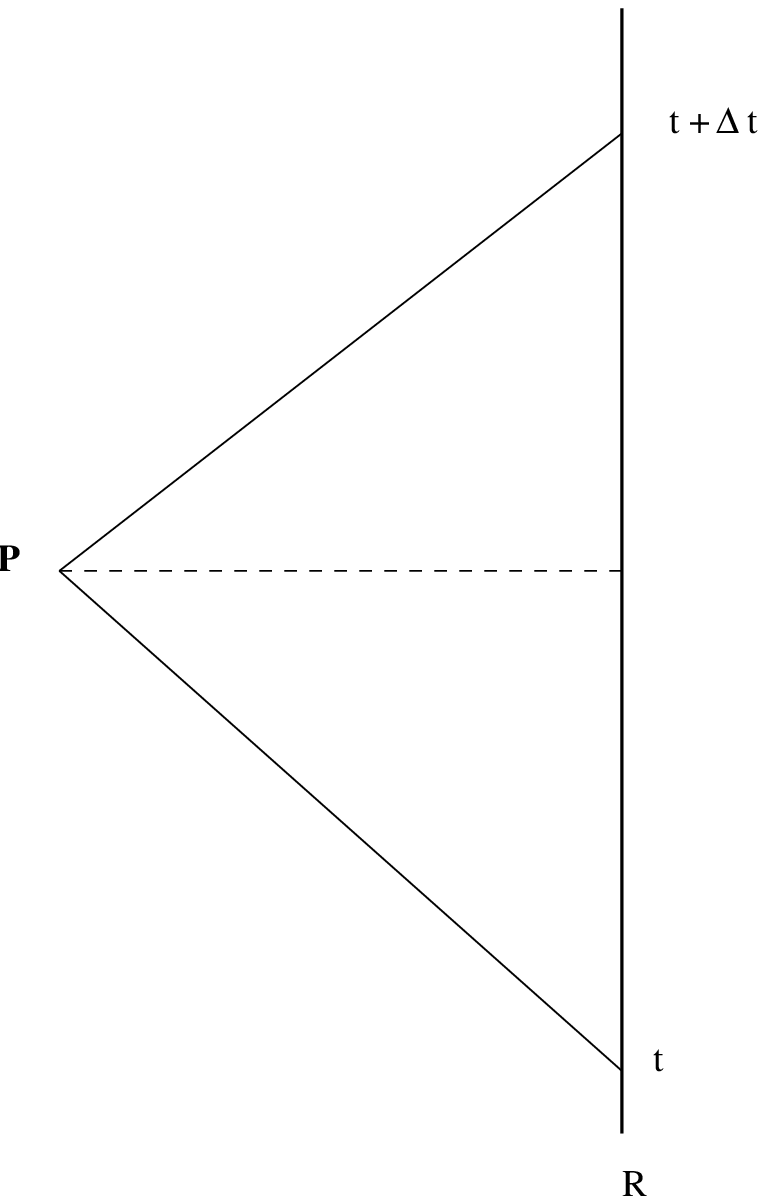}{3.0}

It is useful to be more explicit in describing these states.  To do
so we introduce a particular choice of coordinates.  Suppose that the
metric is asymptotically flat; we base the coordinates on an
asymptotic observer fixed at large radius $R$ from the black hole.
Let the observer carry a clock.  The coordinates of an arbitrary
event $P$ are defined by radiolocation; see fig. 3.  If an inwardly
directed
light signal emitted from $R$ at time $t$ is reflected from $P$ and
returns to $R$ at time $t+\Delta t$, $P$ is assigned coordinates
$$
(r_*,t) =(R-{\Delta t\over2}, t+{\Delta t\over2})\ .
$$
In the case of a static black hole geometry, this prescription gives
the usual tortoise coordinates.  Notice in particular that the
interior of the black hole is not covered.

\ifig{\Fig\Pen}{The Penrose diagram for an initially extremal black hole,
into which some matter is thrown, and which subsequently
evaporates back to extremality.}{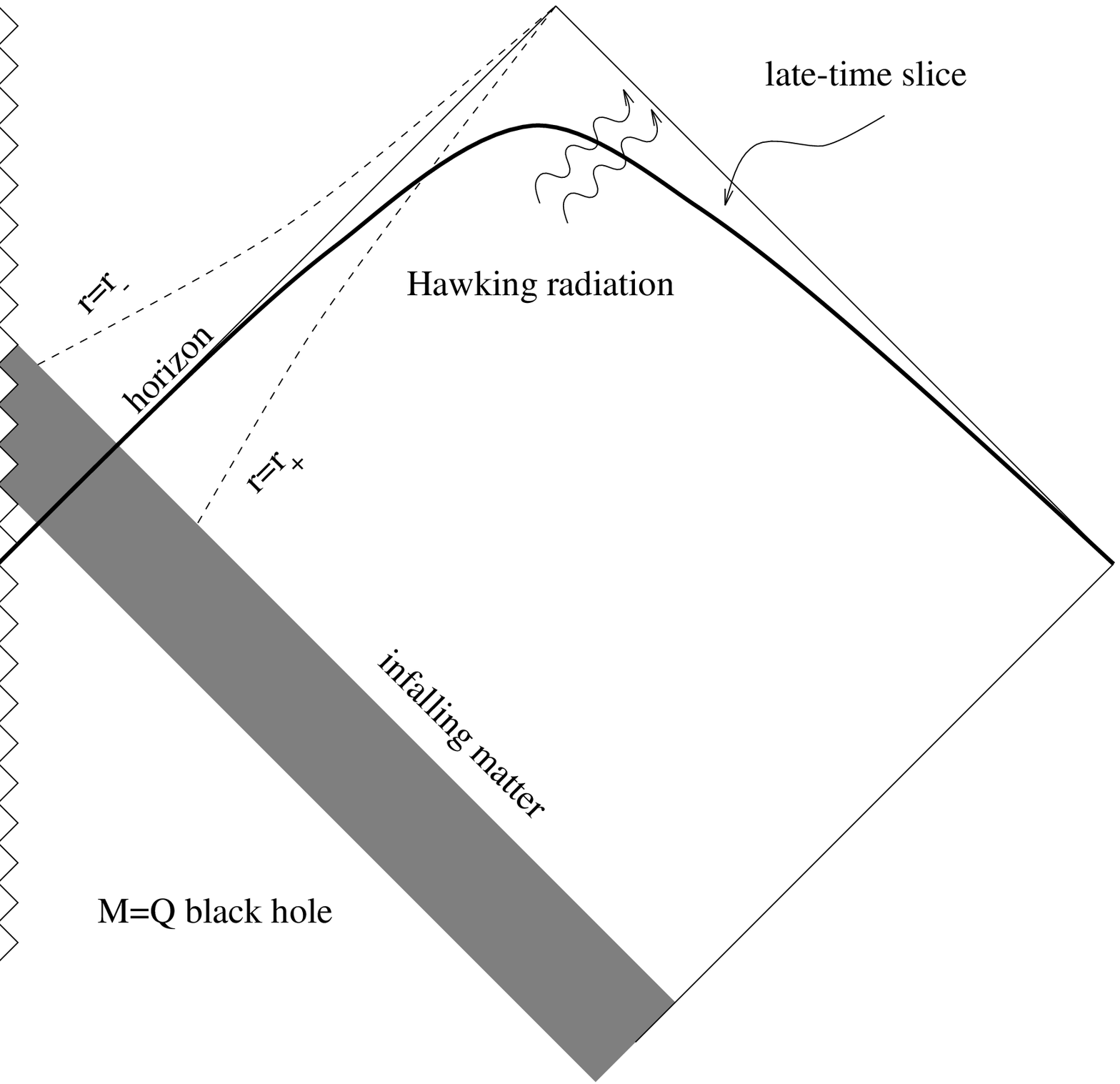}{3.0}

The states of the black hole can be described using data
specified on these time slices.  The time scale for evaporation of a
black hole to extremality\refs{\StTr} is $\tau\sim Q^3$.
Consider the configuration on a slice in the
far future, long after the mass excess $M-Q$ has been radiated past radius
$R$.  Thus in terms of the data on the slice inside $R$, the
states have energy $Q$ and the external appearance of a black hole.
However, differences between states are seen if one investigates near
the horizon.  The slice crosses the infalling matter, whose different
configurations imply different quantum states of the black hole.
These features are illustrated in figs.~\Pen,5.

\ifigx{\Fig\Slice}{Shown is a schematic description of the state of the
black hole of fig.~4 on a late time slice that stays outside the true
horizon }{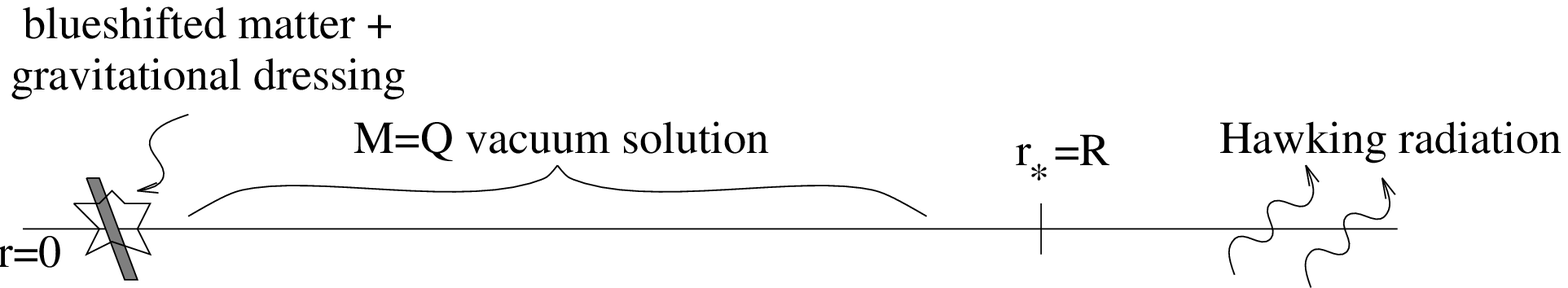}{4.5}

At large times the Hawking radiation turns off and the black hole
must asymptote to a superposition of its exact energy
eigenstates.  Semiclassically these states all have the same
appearance.  The matter distribution asymptotes to $r_*=-\infty$, and
outside the solution should be the extremal vacuum geometry. In this
geometry the proper distance to the matter also becomes infinite:  the
extremal \RN\ black hole has an infinitely long throat.  In the long-time
limit, the only difference between solutions is in the matter configuration
at the end of this throat.

Of course,
the semiclassical description based on our slices eventually fails.  One
way of estimating where this happens is to inquire when observers
traveling on worldlines of fixed $r_*$ see the proper frequencies or
wavenumbers at the Planck scale.  For example, and outgoing s-wave of the
Hawking radiation for a massless field behaves like
$$
e^{-i\omega(t-\rst)}\ ;
$$
at infinity the typical frequency is $\omega\sim T_H$.  Planck physics
becomes relevant at the radius where the proper frequency  becomes
planckian; if $n^\mu$ is the unit normal to the time slices, this occurs
where $n^\mu k_\mu \sim \Mpl$.  Likewise, if the infalling matter is
followed in, the description fails when it becomes planckian.
Notice that while one ordinarily
expects to have a description of the infalling state that does not require
Planck scale physics if one uses
the frame of the infalling observer, it is the
translation of this description to the frame of the outside observer that
requires
planckian physics.   Therefore it seems that the differences
between the infinite number of states are not discernible
without a full theory of quantum gravity.

As stated above, there are several distinct types of black hole,
depending on whether one began
with a truly extremal black hole, with one end of a wormhole, or with a
non-extremal black hole.  However, in each case the final state of the
Hawking process is a solution with $M=Q$, and in the semiclassical
description these ground state solutions
differ only in the configuration at $\rst=-\infty.$  It is not known if
this statement is modified in the full quantum theory.

In our later discussion black holes that are thermally excited will
play a central role.  Suppose we take an $M=Q$ black hole, and
place it in a box of blackbody radiation.  The black hole will then
absorb radiation until the accretion rate and the Hawking radiation
rate match; this should happen when the temperature of the radiation
and black hole are equal.  For neutral black holes this equilibrium
can be arranged to be stable, despite the negative specific heat, by
taking the radius of the box to be sufficiently small, $r\sim
M^{5/3}$.  For charged black holes sufficiently close to extremality
the specific heat is positive, so this is even easier to achieve.
If we compute the partition function, Tr$e^{-\beta H}$, for states in the
box, the infinite
number of ground states will contribute and the partition function
will therefore diverge.

As first pointed out by Gibbons and Hawking\GiHa, an elegant path integral
derivation of the partition function also exists.  The evolution
operator $e^{-\beta H}$ can be turned into a euclidean path integral
by the usual steps, and the trace corresponds to the periodic
identification.  Thus one is instructed to sum over asymptotically
flat geometries with period $\beta$ at infinity.  One ordinarily
assumes that these geometries should be regular in the vicinity of
the horizon.  However, in accord with the above arguments, doing so
would discard the contribution of the infinite number of states:
these correspond to configurations that do not behave smoothly at
the horizon.  Thus it seems that the instruction to sum over regular
geometries only captures a finite subset of the states,  and does not
give the correct result for Tr$e^{-\beta H}$.  The infinite states
only appear to be accounted for if one allows singular behavior in
the vicinity of the horizon.\foot{It seems
quite likely to the present author that the contribution of the
infinite number of states and the conformal factor problem are
closely connected.  Indeed, the divergent integral over the
conformal factor quite likely is connected to the required infinity
in the partition function, and the unstable behavior seems connected
to the irregularity of the geometry. (A possibly related comment has been
made by Carlip and Teitelboim\CaTi.)}  Although the trace must
contain an infinite factor from the infinite number of states, the
prescription for a detailed calculation
cannot be given in the absence of a
quantum theory of gravity.

It should be emphasized that the infinite number of states contributing to
the trace can be explicitly counted.  For example, one could imagine
forming black holes from diffuse collapsing matter.  The initial state of
this matter can be taken to be in finite volume, and can be specified in
the presence of a short distance cutoff.  The number of such states that
collapse to form a black hole is therefore enumerable and finite.  If one
assumes unitary evolution (and in particular conservation of the norm of a
state) and that information does
not escape from black holes, then the infinite volume limit
gives an infinite number of final states
containing a black hole and in which the  only
difference is in the internal state of the black hole.
This infinity in the number of states
should also appear in the quantity Tr$e^{-\beta H}$.  Other
authors\refs{\SuUg,\FPST}  have advocated calculations that give a finite
result for the latter quantity.  It would seem that this is only possible
either if these calculations are not including all black hole states or if
our assumptions are wrong and black holes only have finitely many states.

\newsec{Pair production via tunneling}

A good starting point for the description of Schwinger production is the
Wheeler-deWitt equation, or its completion in a full theory of quantum
gravity.  This equation acts on wavefunctionals
$$
\Psi[\threeg,f(x),A(x),T]
$$
of the three geometry, the matter fields $f$, the gauge field $A$, and
asymptotic time $T$.
The solutions of this equation are given by the lorentzian path integral.
Where the semiclassical approximation is valid and in classically allowed
domains, leading order solutions of
this equation are simply given in terms of classical solutions of the
coupled Einstein-Maxwell-matter equations according to the standard WKB
formalism.

One can likewise consider tunneling through classical forbidden
regions, as in the Schwinger process.
Again where the semiclassical approximation is valid, the leading
semiclassical wavefunctions are given by classical solutions, in this case of
the euclidean continuation of the equations of motion.\foot{The connection
between real-time tunneling and euclidean solutions for other field
theories is made explicit in \refs{\tunneling,\BiCh}.}  This gives the
tunneling rate to a configuration that is a classical turning point.  The
system can also tunnel to nearby configurations via paths near the
euclidean solution.  The contribution to the tunneling rate
of these nearby configurations
is well-approximated
by the usual fluctuation
determinant\refs{\BiCh}, as in standard instanton calculations.
Alternatively, these results can be obtained directly from the euclidean
functional integral\refs{\CoUses}.

There are two types of euclidean solutions describing pair production of
black holes.\foot{For a more complete discussion see \refs{\DGGH}.}  The
first\refs{\GaSt,\GGS} describes production of a pair of oppositely-charged
black holes connected by a Wheeler-wormhole throat.  The black holes are
consequently above extremality.  The second\refs{\Gibb,\DGGH} describes
production of an oppositely charged pair of extremal black holes that are
not connected.  For simplicity we will henceforth focus on creation of
magnetically-charged black holes in a magnetic field.
Then both of these solutions asymptote to the Melvin
universe\refs{\Melv}, which is the closest approximation to a uniform
magnetic field in general relativity.

Necessary conditions for
validity of the semiclassical approximation are that $Q\gg 1$
(super-planckian black holes) and $QB\ll 1$  (weak magnetic fields).
The
leading semiclassical tunneling rate is given by the action,
\eqn\scrate{\Gamma\sim e^{-S}\sim e^{-\pi m^2/QB}\ .}
However, this estimate clearly misses the contributions to the
tunneling rate of the infinite
number of states.  The instanton describes
tunneling to the classical turning point, which is a pair of vacuum \RN\
black holes.  However, nearby configurations, reached by nearby paths, have
non-trivial matter and gravitational excitations.  Following our preceding
discussion, these should include the infinite number of states of the black
hole.  Their contributions are therefore included to linear order
by the fluctuation
determinant around the instanton, or including interactions,
by doing the full euclidean functional integral in the vicinity of
the saddle point.

Even the fluctuation determinant is difficult to compute directly.
However, it is closely related to the result of another calculation, namely
that of the thermal partition function for a \RN\ black hole.

To see this, let us give a more detailed description of the
Ernst instanton solutions\refs{\Ernst}.  For simplicity focus on the
magnetic case.  They are:
\eqn\dernst{
\eqalign{
ds^2&=(x-y)^{-2}A^{-2}\Lambda^{2}
\left[\left\{G(y)dt^2-G^{-1}(y)dy^2\right\}
+G^{-1}(x)dx^2\right]\cr &\qquad +
(x-y)^{-2}A^{-2}\Lambda^{-2}G(x) d\varphi^2\cr
A_\varphi&=-{2\over B\Lambda}\left[1+{1\over 2}Bqx\right]
+k\ ,\cr
}
}
where the functions $\Lambda \equiv\Lambda(x,y)$  and $G(\xi)$ are given by
\eqn\fns{\eqalign{
&\Lambda=\left[1+{1\over 2}Bqx\right]^2+{B^2\over 4A^2(x-y)^2}
G(x) \cr
&G(\xi)=(1-\xi^2-r_+A\xi^3)(1+r_-A\xi) \cr
}}
and $q$ is given by
\eqn\qdef{q^2 = r_+ r_-\ . }
$A$ and $B$ are parameters, and the constant $k$
in the expression for the gauge field is introduced
so that the Dirac string of the magnetic field of a black hole
is confined to one axis.  Finally, it is useful to factorize $G$,
\eqn\Grew{G= -r_+r_- A^2(\xi-\xi_1)(\xi-\xi_2)(\xi-\xi_3)(\xi-\xi_4)\ ,}
with
\eqn\xidef{\xi_1=-{1\over r_- A}}
and $\xi_2$, $\xi_2$,  $\xi_2$ the ordered roots of the remaining
expression.  For small acceleration, $r_+A\ll 1$, the zeroes have
expansions
\eqn\xiexp{\eqalign{\xi_2 &= -{1\over r_+ A} + r_+ A + \cdots\cr
\xi_3&= -1 -{r_+ A\over 2} + \cdots\cr
\xi_4&= 1-{r_+A\over 2} +\cdots .}}

The solution \dernst\ describes a pair of black holes with opposite
magnetic charge in a background magnetic field.
The independent parameters of this solution are $r_\pm$, $A$, and
$B$, to be thought of (roughly) as the inner and outer horizon radii, the
acceleration, and the magnetic field strength.  For general
parameters the solution \dernst\ is not regular.  In particular,
if the
acceleration is not related to the charge, mass, and magnetic field, then
there will be a physical string singularity attaching the two black holes.
With these parameters matched, the lorentzian geometry is regular outside the
horizons, and it can readily be shown that the black holes follow
approximately hyperbolic trajectories corresponding to uniform
acceleration.  The time $t$ used in
\dernst\ corresponds to Rindler time asymptotically far from the black
hole, as can be shown by investigating the limit $x\rightarrow y$.\foot{For
a more detailed description of
the features of
this solution, see for example \refs{\DGGH}.}

The instanton for pair production follows from substituting $\tau=it$.  Now
there is another condition that must be imposed on the parameters to have a
regular solution.  To see this, note that the point $y=\xi_3$ corresponds
to the acceleration horizon and regularity there requires a specific
periodicity for euclidean Rindler time $\tau$ as in standard treatments of
Rindler space.  However, $y=\xi_2$ corresponds to the black hole horizon,
and a periodic identification of $\tau$ is also required there as in
standard treatments of the black hole.  Equating these periods gives a
second
relation between the parameters.  In a sense, this is a condition
matching the acceleration and Hawking temperatures so that
the black hole can be thought of as being in static equilibrium with the
acceleration radiation.

Consider the case of small acceleration.  There are actually two solutions
to the temperature-matching condition.  The first\refs{\GaSt}
is if the black hole is
taken to be slightly above extremality in order to raise the temperature
enough above zero to match the acceleration temperature.  When matched onto
lorentzian solutions, these instantons are seen to create pairs of
non-extremal black holes connected by a Wheeler wormhole.
The second\refs{\Gibb} is at
first sight surprising:  for truly extremal black holes, the horizon is at
infinite proper distance and so any periodicity is allowed.  This latter
case corresponds to the limit $\xi_1=\xi_2$ and pair produces extremal
black holes.

In a quantum treatment the solutions will receive corrections from
the backreaction of the Hawking/acceleration radiation on the
geometry.  Equilibrium with thermal acceleration radiation is quite
analogous to equilibrium with a thermal bath.  In particular, the
extremal case is likely no longer a solution as the black hole is
raised above extremality.  Therefore we focus on the non-extremal
wormhole solutions.

Possible contributions of the infinite states arise in computing
the functional
integral over configurations near the instanton. This is hard.  However, let
us investigate the instanton  in the throat region, where in accord
with our earlier discussion the infinite states are expected to be
located if they are present.

The vicinity of the black hole corresponds to $y\rightarrow \xi_2$ in
\dernst.
Using the periodicity-matching condition,
\eqn\pmatch{\xi_1-\xi_2-\xi_3+\xi_4=0}
and the expansions \xiexp, the metric takes the form
\eqn\assoln{ds^2 \rightarrow q^2\left[ -\sinh^2w dt^2 + dw^2 + d\Omega_2^2
\right]\ }
after a change of variables.
This agrees exactly with the form of the free near-extremal
\RN\ solution near the
horizon, as can be seen from the substitution
\eqn\rnsubs{{r-r_+\over r_+-r_-} = {1\over 2} (\cosh w -1)\ }
and rescaling $t$.
Subleading corrections to this expression vanish in the limit
$qB\rightarrow0$ and $w\rightarrow0$. In particular,
the leading correction
to $g_{tt}$ is of the form $qB(\cosh w -1)\sim qBw^2$.
These corrections are small and furthermore
do not shift the location of the horizon or qualitatively change the
form of the solution in the vicinity of the black hole.  The corrections do
become substantial, however, when $w\sim -\ln(qB)$, where the transition to
the asymptotic solution takes place.  The length of the black hole throat
is therefore $l\sim -q\ln(qB)$.  The corrections are
exponentially small in the length of the black hole throat.  Finally, note
that to leading order in the $qB$ expansion, the parameter $q$ and the
physical charge $Q$ are equal.

The solution will also receive corrections from the backreaction of the
Hawking/acceleration radiation.  Since all of the effects of the
acceleration, with the exception of the thermal fluxes, die near the
horizon, the backreaction-corrected solution should be of the same form as
that of a free black hole in equilibrium with radiation, plus small
corrections.  Detailed descriptions of such
solutions has not been given, although
backreaction-corrected solutions for extremal \RN\ black holes
without the thermal flux have been investigated in \refs{\Triv} and
dilatonic black holes in equilibrium with an inward flux were found
numerically in \refs{\BGHS,\SuTh}.  Outside the black hole
these are expected to preserve the
general form of the near-extremal solution.  It should be noted that as in
the free case, there are an infinite number of solutions which in the far
future differ only in their state at the horizon.

\ifigx{\Fig\Inst}{A picture of the euclidean instanton solution, for small
$QB$.  Below the dotted line, the solution is nearly identical to the
euclidean \RN\ solution.  The contribution of the infinite number of states
is expected to arise from configurations that rapidly oscillate near the
would-be horizon.}{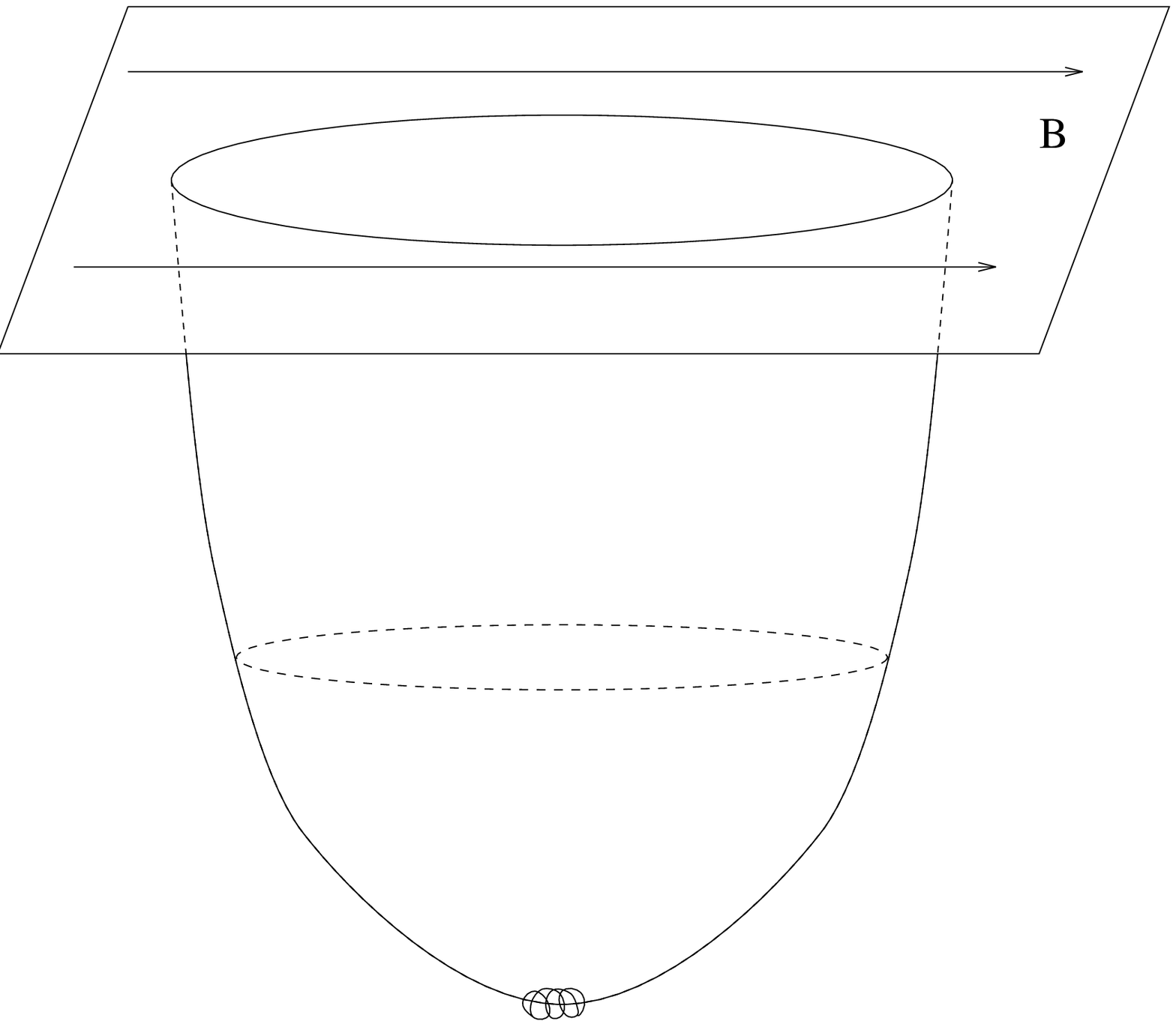}{2.5}

In pair creation, these states are accounted for in the euclidean
functional integral about the instanton.
Once again, we do not know how to evaluate this
integral without understanding quantum gravity.  However, we have just
argued that in the throat region, for
\eqn\thrl{w\ll -\ln(qB)\ ,}
the solution is identical to that of a free black hole in equilibrium with
thermal radiation, up to small corrections.  Indeed, a
quick check shows that the
local temperature at the end of the black hole throat is $T\sim B$, the
expected value from the acceleration radiation.  Although the semiclassical
approximation fails, the contributions to the functional integral from
Planck-scale dynamics should be essentially the same in either case.
Indeed, using the composition
property of the functional integral, it can be split along the dotted line
in fig.~6.  The contribution from the bottom of the cup should be
approximately the {\it same} as that from
corresponding region in the computation of the
euclidean functional integral for free
black hole in
contact with a thermal bath.
As explained in section three, we also can't calculate the latter
functional integral, but we know it gives the partition function.
Thus
the production rate contains a factor of the form
\eqn\bhpart{
{\rm Tr} e^{-\beta H}\ .}
There will of course be differences between this quantity and the
functional integral for the instanton, arising from differences outside the
throat region.  However, if the black hole has
infinitely many states down the throat, there should be contributions of the
form \bhpart\ from these states to the pair creation rate.
Note finally that this factor corresponds to the factor found in the
low-energy effective
calculation of the rate, \Prate.

To summarize these arguments, although the calculation of the functional
integral cannot be done without using details of quantum gravity, the
calculation should be the same as that for the throat contribution
in Tr exp$\left\{-\beta H\right\}$.
If a Reissner-Nordstrom black holes of charge $Q$ has
infinite numbers of nearly-degenerate ground states, there is a
corresponding infinity in both expressions; the pair creation rate is
infinite.

Although the contributions of the infinite states come into the instanton
calculation through singular geometries, note that there are also smooth
geometries that contribute to the functional integral:  these are precisely
the original Wheeler wormhole configurations, with the `internal' states
unexcited.  In accordance with the arguments of
\refs{\StTr,\CILAR,\Yi} it is quite plausible that pair creation for
these regular Wheeler wormholes is in fact finite because they
are rather special states. More general states are found by throwing matter
into a Wheeler wormhole and then letting it evolve back to equilibrium with
the radiation.

Finally, an interesting question is what is the typical state of the Hawking
radiation for the pair created black holes.  Since
the euclidean section of
the instanton closely approximates the
euclidean section of the unaccelerated black hole away from the horizon,
the Green functions for excitations are computed according to
the Hartle-Hawking prescription.  This ensures that the state produced is
essentially the Hartle-Hawking state\refs{\CILAR,\Yi}.

\newsec{Conclusions}

If we assume the validity of quantum mechanics and also that information is
not returned in Hawking radiation, this seems to inevitably lead to the
statement that the Hawking process leaves behind an infinite number of
``remnant'' states in the evaporation of a black hole.  We have argued that
if this is the case, there is no obvious mechanism for suppressing the
resulting infinite pair-creation rate for \RN\ black holes.  This appears
to be a catastrophe.

There are several ways to attempt to escape this conclusion.  Let us
consider them in turn.

One possibility is that extremal black holes don't exist as true ground
states of any physical theory.
A charged black hole itself sheds charge by Schwinger production, at a
rate\refs{\Gibbsp}
\eqn\Charged{{dQ\over dt} \simeq {e^4 Q^3\over r_+} e^{-\pi m^2 r_+^2/eQ}}
for quanta of mass $m$ and charge $e$.  This can for example
be compared to rate of
change of the mass through Hawking emission.
In the electric case,
black holes will rapidly
discharge through
electron emission unless $M\roughly> 10^5 M_\odot$.  The situation is
improved in the magnetic case.  If one for example considers a grand
unified theory, production of magnetic monopoles by extremal black holes
is highly suppressed for
\eqn\Monbd{M\gg g/M_{\rm mon}^2\ ,}
a much more reasonable constraint.\foot{Note that the instability of
\refs{\LNW} is also absent for sufficiently large charge.}  Furthermore, for
\eqn\Mforbid{M\roughly>gQ/M_{\rm mon}}
monopole emission is forbidden.  We can therefore easily create
a charged black hole with an infinite number of internal states by
beginning with a black hole satisfying \Mforbid\ and feeding it
information
and energy at a sufficient rate to balance the Hawking radiation for as
long as we please.  If it then Hawking decays to extremality, one
finds an infinite number of species of metastable extremal black holes.  In
\refs{\CBHR} it was argued that for remnant decay lifetimes larger
than the Schwinger time,
\eqn\Stime{\tau_S\sim l_S \sim M/QB\ ,}
finite decay rates do
not substantially affect pair production.  The exponential suppression of
\Charged\ makes
this easy to achieve for moderate $Q$.

If one instead worked in a theory with no GUT monopoles, it is quite
possible that magnetic black holes could be pair produced as Wheeler
wormholes and then be absolutely stable to Schwinger emission.  In any case,
even if discharge instability were to make pair creation of \RN\ black
holes finite, this would just shift the infinite production problem into
the neutral remnant sector.

A second attempted out is to appeal that Schwinger production of black
holes requires a very strong field that is uniform over extremely large
scales.  Indeed, for true Schwinger production the field should be uniform
over at least the magnetic length of \Stime,
\eqn\Uniform{L\roughly> {10^{20} cm\over B(tesla)}\ ,}
which is not likely to be realized.
However, for much weaker fields that are
non-uniform, one also expects there to be a finite but minuscule
production rate for each species
as long as there is sufficient energy available to make a
pair of black holes, $E\gg 2\Mpl$.  Although this rate has enormous
supressions due to form factors, etc.,  these are overcompensated by the
overall infinite number due to the infinite number of remnant species.  If
Schwinger production is not finite, it's probably not possible for these
rates to be finite either.

A third possible escape is that although the difference between the instanton
geometry and the geometry of the euclidean asymptotically flat black hole
vanishes far down the throat, this small difference conspires with Planck
scale physics to make the calculation of the fluctuation determinant
differ from that of the partition function by an infinite factor.  In light
of the fact that without this the calculation seems to be giving one a
result in agreement with effective arguments, and in accord with the
implications of crossing symmetry, this seems unlikely.

A fourth possibility is that despite the fact that black holes have infinite
number of states, there is a prescription to
compute $Tr e^{-\beta H}$ for a black hole that gives a finite answer,
and furthermore there is a reason that this is the correct prescription to
use in calculating the production rate.
One proposal is that the infinity can be absorbed
through
renormalization of Newton's constant\refs{\SuUg,\FPST}.\foot{I thank
A.~Strominger for conversations on this issue.}  However, this seems
unlikely to work as Newton's constant should be renormalized to give
correct low-energy gravitational scattering amplitudes at low energies.
Once this renormalization has been done, it is still apparently true that
black holes have infinite numbers of states, and thus the trace over black
hole states should still have a non-subtracted infinity.

A fifth possibility is that $Tr e^{-\beta H}$ is finite because
Hawking was right:  information is lost in
quantum gravity, and this information loss causes black holes to have only
a finite number of states.  However the serious conflicts with energy
conservation\refs{\BPS,\Sred,\CILAR,\Trieste} that arise from this possibility
remain; there is no known effective description of local
information loss that conserves energy.  This is a major problem.

The final possibility is that black holes have a finite number of states
and information is conserved: it is emitted in the Hawking process.
Despite the fact that this
possibility has recently been vigorously pursued\refs{\tHoo\STU-\Suss},
there is as of yet little evidence of a concrete mechanism for string
non-locality or other physics
to imprint information on the Hawking radiation.  Nonetheless,
given the results of this paper it is possible that the only way to avoid
infinite production is if the information indeed comes out in the Hawking
radiation.

\bigskip\bigskip\centerline{{\bf Acknowledgements}}\nobreak

I wish to thank G. Horowitz, M. Srednicki, and A. Strominger
for helpful discussions.  This work was supported in part by
DOE grant DOE-91ER40618 and
by NSF PYI grant PHY-9157463.

\listrefs

\end